# An Efficient Hidden Markov Model for Offline Handwritten Numeral Recognition

[1]B S Saritha, [2]S Hemanth
[1]M. Tech, IV Semester, Department of CSE, CMRIT
Software Architect, IBM, Bangalore, India. sarithamys@yahoo.com
[2]Assistant Professor, Department of Computer Science and Engineering,
RNSIT, Bangalore, India, hemanth.shantha@gmail.com

## Abstract

**Traditionally, the performance of ocr algorithms and systems is based on the recognition of isolated characters. When a system classifies an individual character, its output is typically a character label or a reject marker that corresponds to an unrecognized character. By comparing output labels with the correct labels, the number of correct recognition, substitution errors misrecognized characters, and rejects unrecognized characters are determined. Nowadays, although recognition of printed isolated characters is performed with high accuracy, recognition of handwritten characters still remains an open problem in the research arena. The ability to identify machine printed characters in an automated or a semi automated manner has obvious applications in numerous fields. Since creating an algorithm with a one hundred percent correct recognition rate is quite probably impossible in our world of noise and different font styles, it is important to design character recognition algorithms with these failures in mind so that when mistakes are inevitably made, they will at least be understandable and predictable to the person working with the program.**

**Index Terms –** *Hidden Markov Model, Optical Character Recognition, Offline Handwritten*

## 1. INTRODUCTION

Since the advent of digital computers, there has been a constant effort to expand the domain of computer applications. Some of the motivations for this effort come from important practical needs to find more efficient ways of doing things and some from the sheer challenge of building or programming a machine to do things that machines have never done before. Both of these motivations can be classified in the area of artificial intelligence, called machine perception. Character recognition is an important topic in pattern recognition, which is the backbone of machine perception. Invention of scanners and improvement in scanner technology has enriched the applications of automated character recognition.

Handwritten data can be generated in two different ways, on-line and off-line. In the former, the data were captured during the writing process by a special device Stylus pen on a tablet. In the latter, the data are captured by scanner after the writing process is over. In this case, the recognition of off-line handwriting is more complex than the on-line case due to the presence of noise in the image acquisition process and the loss of temporal information such as the writing sequence and the velocity.

This information is very helpful in the recognition process. Off-line and on-line recognition systems are also discriminated by the applications they are devoted to.

The off-line recognition is dedicated to bank check processing, mail sorting, reading of routine commercial forms. The on-line recognition is mainly dedicated to pen computing industry and security domains such as signature verification and author authentication. This work is limited to the off-line recognition.

For more than 35 years researchers have been working on an OCR system based on shape that has the accuracy of human vision. In order to facilitate future progress in document analysis, there is a need for a number of scanned document datasets, each representing a different class of documents, text, engineering drawings, addressed forms, handwritten manuscripts etc.

## 2. STAGES

The OCR system for hand written numeral used in this work has the following stages.

### 2.1 Preprocessing

The raw data is subjected to a number of preliminary processing steps to make it usable in the descriptive stages of character analysis. Pre-processing aims to produce data that are easy for the OCR systems to operate accurately. The main objectives of pre-processing are:

- Noise Reduction
- Stroke Width Normalization

- Segmentation
- Thinning

In the preprocessor stage smoothing and normalization is performed. Filling and thinning to eliminate noise, removal of isolated pixels and skeletonization is done in smoothing.

Writing styles differ from person to person, hence there is a large variation in the character / numerals and also writing style of a person varies with times. Hence in a character / numeral recognition it requires a large database to account for the different styles of writing. A dataset of about 40 samples of each numeral has been created. Each sample is scaled to fit into a 64x64 window and then thinning algorithm is applied to obtain the thinned / skeletonized image, which is used in feature extraction.

The thin-line representation of certain elongated patterns, for example characters, would be closer to the human conception of these patterns; therefore, they permit a simpler structural analysis and more intuitive design of recognition algorithms. In more practical terms, thin-line representations of elongated patterns would be more amenable to extraction of features such as end points, junction points and connections among the components whereas, some algorithms used in pattern recognition tasks also require one pixel wide line as input. For thinning algorithm to be really effective, it should ideally compress data, retain significant features of the pattern and eliminate local noise without introducing distortions of its own. Thinning is normally applied to binary images, and produces another binary image as output. The deletion of a pixel ‘p’ would depend on the configuration of pixels in a local neighborhood containing ‘p’.

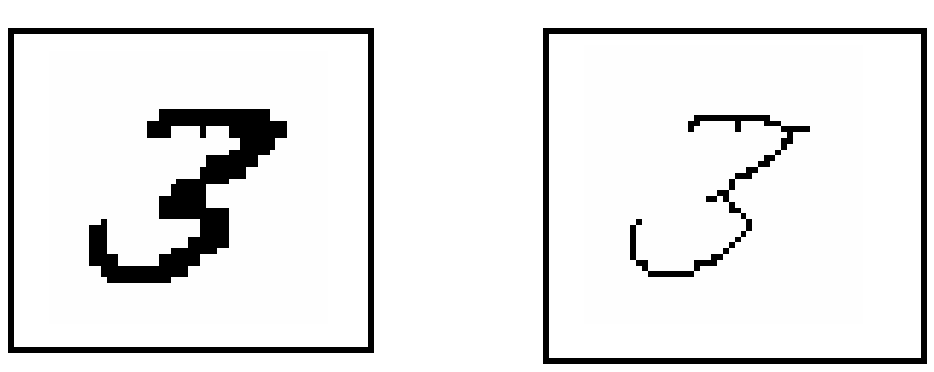

Figure 1**:** Sample of Thinned Character

### 2.2 Feature Extraction

In feature extraction stage each character is represented as a feature vector, which becomes its identity. The major goal of feature extraction is to extract a set of features, which maximizes the recognition rate with the least amount of elements. Feature extraction methods are based on 3 types of features:

- Statistical
- Structural
- Global transformations and moments

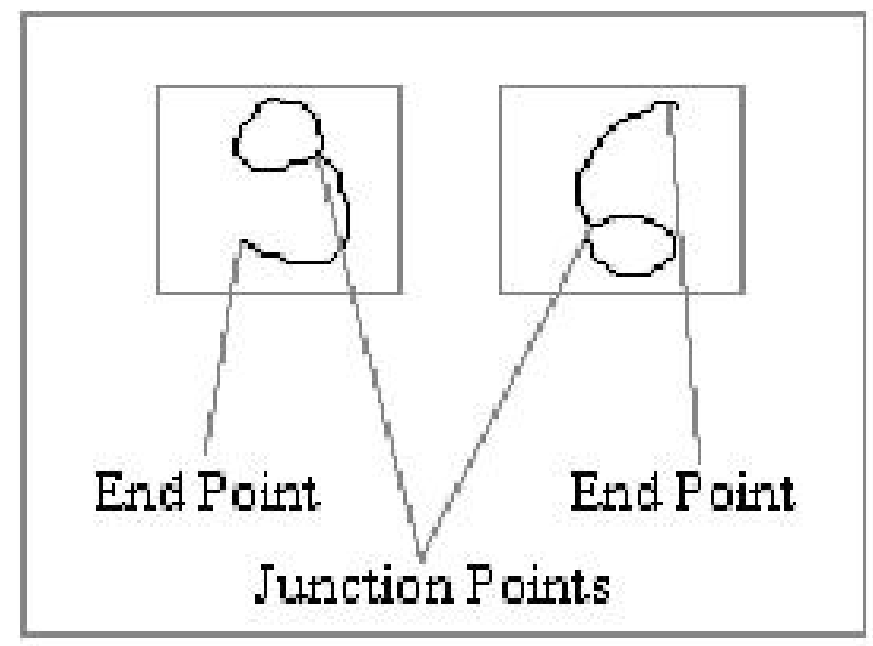

Figure 2**:** Typical Features of Character.

Characters can be represented by structural features with high tolerance to distortions and style variations. This type of representation may also encode some knowledge about the structure of the object or may provide some knowledge as to what sort of components make up that object.

Structural features are based on topological and geometrical properties of the character, such as aspect ratio, cross points, loops, branch points, strokes and their directions, inflection between two points, horizontal curves at top or bottom, etc.

Feature extraction forms the backbone of the recognition process. Features vary from global analysis. Structural analysis includes features such as junction points, end points, loops etc. These features are unique to each numeral. The junction points and end points form the characteristic points. For example, numeral 6 has one junction point and one end point and numeral 8 may have single junction point or two junction points. It totally depends on how the numeral is written.

Extraction of characteristic points can be of:

- Transition function
- End points
- Junction points
- Conic’s representation
- Directional codes

**Transition Function:** Each pixel Pi of a skeleton, resulting from the thinning process, has a transition function T (Pi) associated to it. T(Pi) represents the connectivity between Pi and its eight neighbours. T(Pi) is defined as the number of transitions of 0 (white) to (black) when the eight neighbours of Pi ( i.e. P1, P2, …, P8) are traced in a clockwise direction.

**End Points:**

An end point is a pixel Pi with T(Pi) = 1.

**Junction Points:** A junction point is a pixel Pi with T

(Pi) = 3. After the thinning process, the numeral tracing is ready and we find the junction points and end points.

Using the characteristic points the pats are evaluated using Conic's representation and directional code.

## 3. CLASSIFIER

Once the character has been represented as a feature vector, a classification rule has to be defined in order to classify it into one of the letters. Numerous techniques for off-line handwriting recognition have been investigated based on four general approaches of pattern recognition, such as template matching, statistical techniques, structural techniques and neural networks.

Statistical techniques are concerned with statistical decision functions and a set of optimal criteria, which determine the probability of the observed pattern belonging to a certain class. Several popular handwriting recognition approaches belonging to this domain are:

- k-Nearest-Neighbour (k-NN)
- Bayesian
- Hidden Markov Model (HMM)
- Fuzzy set reasoning
- Support Vector Machine (SVM)

An HMM is called discrete if the observations are naturally discrete or quantized vectors from a codebook or continuous if these observations are continuous. HMM is been proven to be one of the most powerful tools for modeling speech and a wide variety of other real-world signals. These probabilistic models offer many desirable properties for modeling characters or words. One of the most important properties is the existence of efficient algorithms to automatically train the models without any need of labeling pre-segmented data.

In order to define an HMM completely, the following elements are needed.

- The number of states of the model, N.
- The number of observation symbols in the alphabet, M. If the observations are continuous then M is infinite.
- A set of state transition probabilities

$$\wedge = \{a_{ij}\}$$

$$a_{ij} = p\ \{q_t = i\},\ 1 \leq i, j \leq N,$$

Where, $q_t$ denotes the current state.
Transition probabilities should satisfy the normal stochastic constraints,

$$a_{ij} \geq 0,\ \ 1 \leq i, j \leq N \text{ and}$$

$$\sum_{j=1}^{N} a_{ij} = 1,\ 1 \leq i \leq N$$

A probability distribution in each states,

$$B = \{b_j\ (k)\}$$

$$B_j(k) = p\{o_t = v_k | q_t = j\},\ 1 \leq j \leq N,\ 1 \leq k \leq M$$

Where, $vk$ denotes the $k$th observation symbol in the alphabet and the current parameter vector.

The initial state distribution,

$$\pi = \{\ \pi_i\ \} \text{ where,}$$

$$\pi_i = p\{q_1 = i\ \},\ \ 1 \leq i \leq N$$

Therefore we can use the compact notation

$$\lambda = (\wedge, B, \pi)$$

to denote an HMM with discrete probability distributions, while

$$\lambda = (\wedge, c_{jm}, \mu_{jm}, \Sigma_{jm}, \pi)$$

denotes one with continuous densities.

In this work each numeral is modeled by one HMM. The numeral "1" written is modeled by a small number of states.

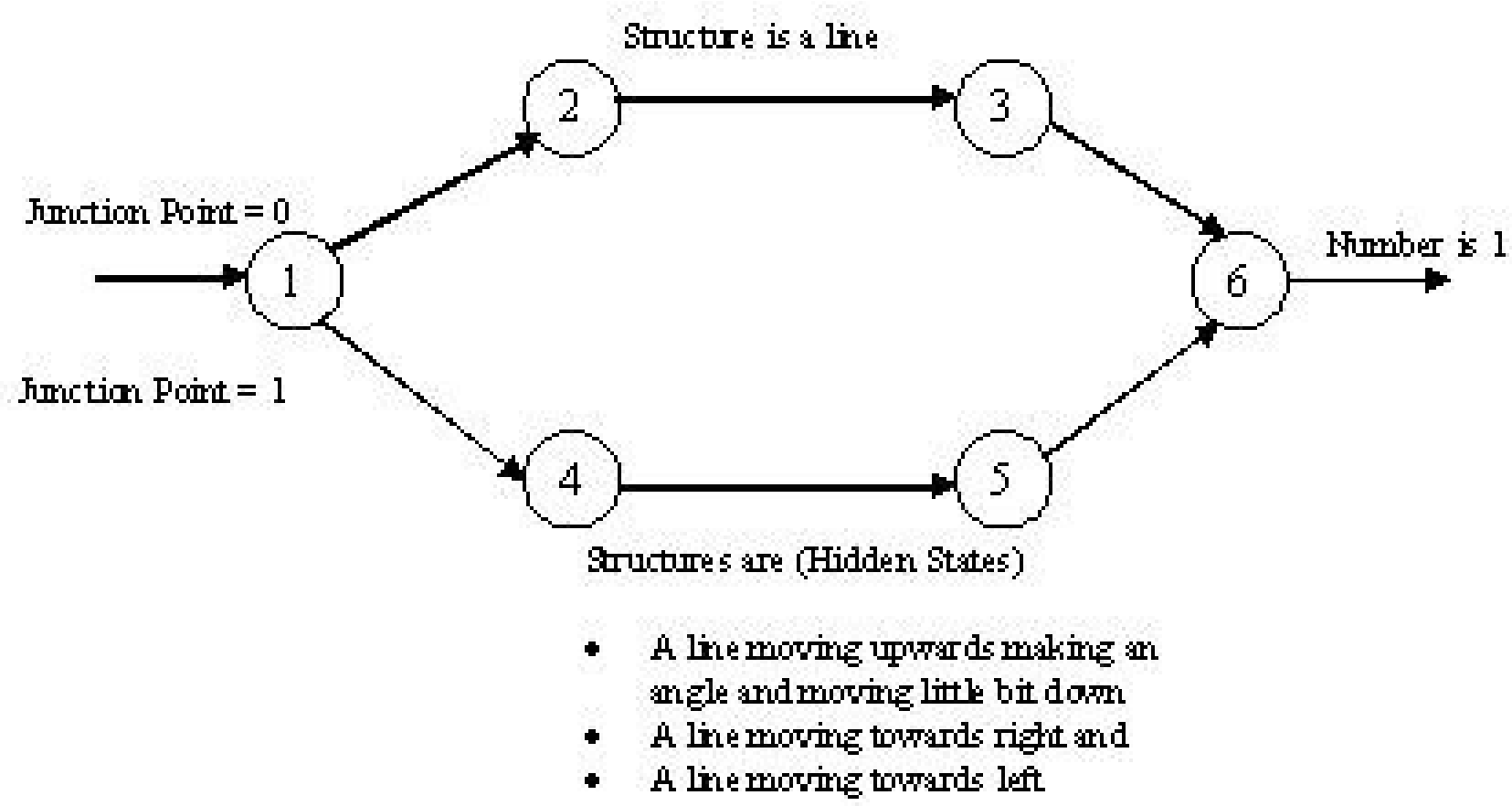

Figure 3. Typical Features of Character

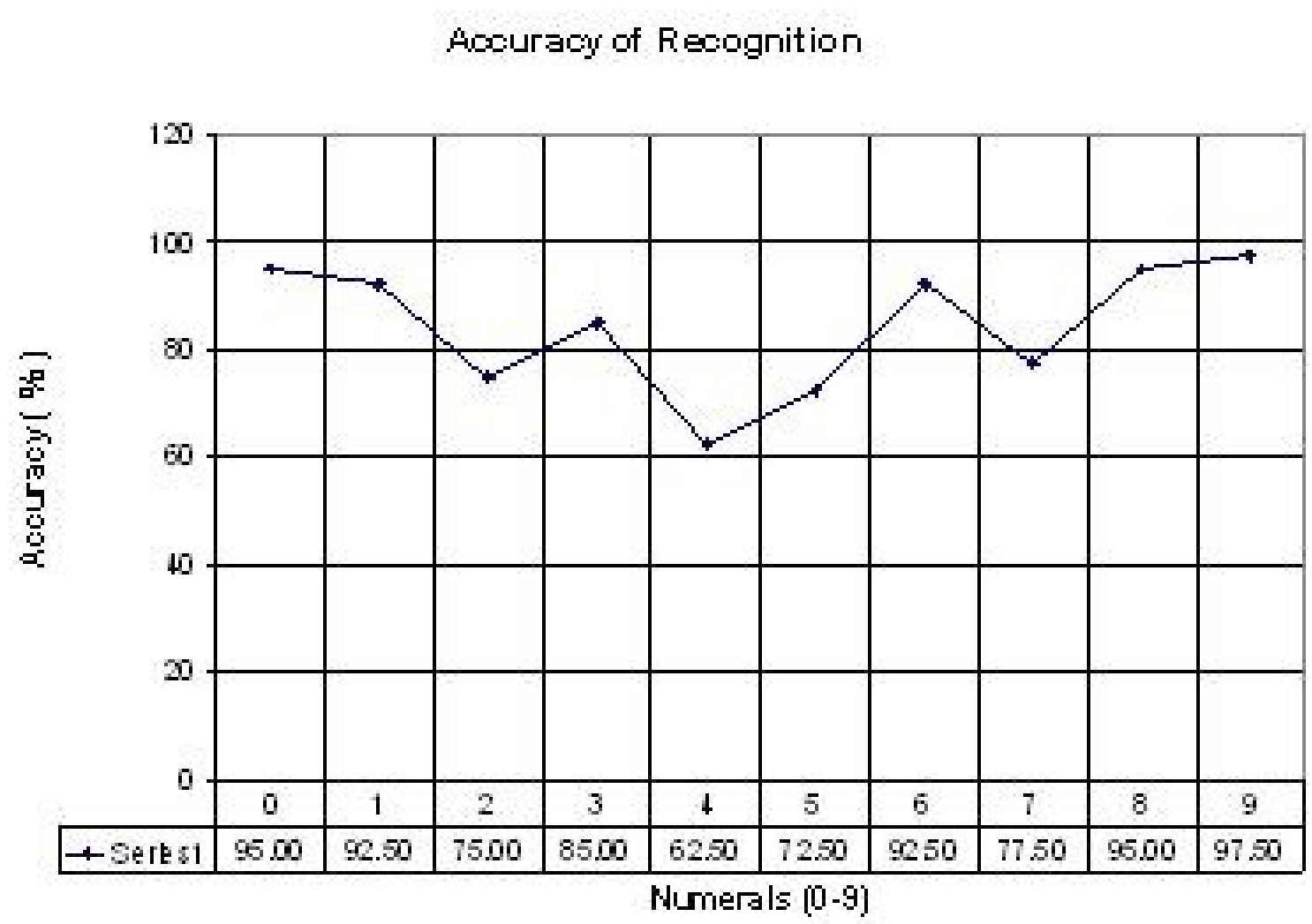

Figure 4 Accuracy of Recognition

## 4. EXPERIMENTAL RESULTS

Standardized downloadable handwritten datasets compiled by National Institute of Standards and Technology (NIST), are used in the experimental work, which is being used by several researchers. Special database 19 contains NIST's entire corpus of training materials for hand printed document and character recognition. It has published hand printed sample forms from 3600 writers, 810,000 character images isolated from their forms. As described earlier, we have used a size normalization step followed by a thinning step before feature extraction.

The model has been tested for about 40 samples of each numeral 0 to 9 and the number of correct recognition is physically recorded. The accuracy of recognition of various numerals is plotted in figure 4. Variation of accuracy of different numerals plotted, which indicates the Number 4 being the least accurate recognizable character and the number 0 being the highest is shown is figure 5.

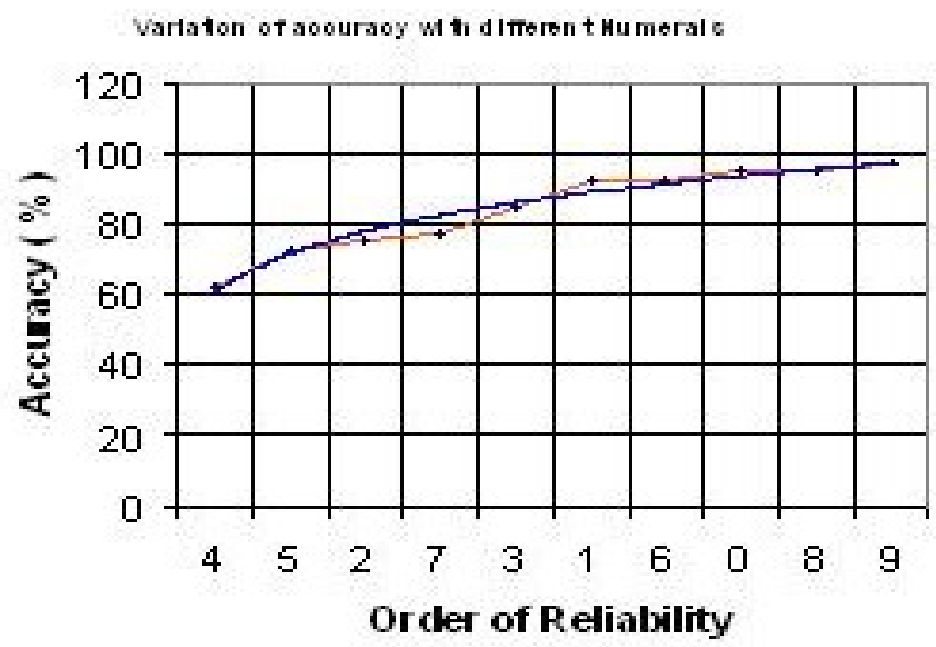

Figure 5 Variation of Accuracy

## 5. CONCLUSION

The objective of handwriting recognition is to have machines, which can read any text with the same recognition accuracy as humans, but at a faster rate. In the present work, a novel method is devised and used for thinning and feature extraction. In the feature extraction, structural features such as junction points, end points and loops have been predominantly used for numerals on a large data set. The features have been used to get HMM model.

The model has been tested considering various numerals from the database and the recognition rate is between 84.5%. It is felt that, finding additional hidden unique features in the numeral can further increase the recognition rate. The thinning algorithm can be further improved by incorporation of heuristic rules to prune the spurious branches in the skeletal image. This will augment to improve the recognition rate of the HMM developed. A combination of HMM along with neural network classifier can also be considered.

## BIOGRAPHY

**B.S.Saritha** graduated from SJ College of Engineering from Mysore. She is fourth semester M.Tech student of CMR Institute of Technology, Bangalore. Her field of interest is image processing and Data Mining.

**Hemanth S** is working as Assistant Professor in RNS Institute of Technology, Bangalore since 2004. He completed his post graduation from SJCE, Mysore. His area of interest are Image Processing , Data Mining and business intelligence.